\documentclass[aps,prd,twocolumn,epsf,nofootinbib,superscriptaddress]{revtex4}

\usepackage{graphicx}
\usepackage{amsmath,amssymb,latexsym}

\begin{document}

\hspace*{130mm}{\tt CAVENDISH-HEP-2009-23}
\vspace{3mm}

\title{Constraining the MSSM with Dark Matter indirect detection data}

\author{Are R.~Raklev}
\affiliation{The Oskar Klein Centre for Cosmoparticle Physics,
Department of Physics, Stockholm University, SE-10691 Stockholm, Sweden}
\author{Martin J.~White}
\affiliation{Cavendish Laboratory, University of Cambridge, JJ Thomson Avenue, Cambridge, CB3 0HE, UK}

\begin{abstract}
Recently, a claim of possible evidence for Dark Matter in data from the Fermi LAT experiment was made by Goodenough and Hooper [8]. We test the Dark Matter properties consistent with their claim in terms of the MSSM by a 24-dimensional parameter scan using nested sampling, excluding all but a very small region of the MSSM. Although this claim is very preliminary, and not made by the Fermi LAT experiment, our scan shows a possible approach for the analysis of future firm evidence from an indirect detection experiment, and its potential for heavily constraining models.
\end{abstract}


\maketitle

The Fermi Gamma Ray Space Telescope (Fermi), launched June 2008 with the Large Area Telescope (LAT) experiment on board~\cite{Atwood:2009ez}, is one of the most powerful current probes in the indirect search for Dark Matter through the detection of its annihilation- or decay-products.  Gamma ray signatures seem particularly generic for Dark Matter models, manifesting as lines from directly produced photons, as a hard spectrum from internal bremsstrahlung, or a softer spectrum from the secondary decays of produced pions. The origin of such photons may be cosmological, from the local halo, from smaller local structures or from the denser inner region of our own galaxy.  Fermi data provides an excellent opportunity for searches in all these categories~\cite{Bergstrom:2007zza,Baltz:2008wd,Ylinen:2008aw,Scott:2009jn,Meurer:2009zz,Vitale:2009zz}.

Following the public release of Fermi data, Goodenough and Hooper~\cite{Goodenough:2009gk} have studied the measured gamma ray flux from the Galactic Centre, defined to be the region $0^\circ
<|l|<3^\circ$, in terms of its angular distribution and energy spectrum. They conclude that the data is well described by a Dark Matter annihilation scenario, assuming a Dark Matter particle mass of $m_{DM}\simeq 25-30$~GeV and a velocity averaged annihilation cross section of $\left< \sigma v \right> \simeq 9\times 10^{-26}$~cm$^3$~s$^{-1}$, with the Dark Matter annihilating dominantly into $b\bar b$ pairs.

At this point we would like to express some doubts as regards the Dark Matter interpretation of the spectra, particularly given that it concerns the notoriously messy environs of the Galactic Centre. For the diffuse galactic background, Goodenough and Hooper assume what is in effect a power-law behavior of the energy spectrum based on data from the angular distribution in the region $3^\circ<|l|<6^\circ$, and its extrapolation into the Galactic Centre as an exponential function. This assumption is clearly vulnerable to features in the diffuse background that have a stronger dependence on the angular distance to the Galactic Centre than the extrapolated function. For a recent criticism, see~\cite{Bringmann:2009ca}. The final word on this matter will naturally rest with the members of the Fermi LAT Collaboration, who are in the best position to understand the difficult backgrounds present.

Despite these doubts, we have found it interesting to speculate how the extracted Dark Matter mass and cross section can be interpreted in terms of the popular Minimal Supersymmetric Standard Model (MSSM). If firm evidence of Dark Matter is indeed found in an indirect detection experiment, then features in the spectrum and the normalization of excesses point to specific Dark Matter masses and annihilation cross sections. It will then be natural to ask how this information can be used to constrain potential models, such as the MSSM. This is the focus of the present Letter.

Naively, one might expect that a low neutralino mass interpretation of the results reported in~\cite{Goodenough:2009gk} is already excluded by collider searches. However, the most commonly quoted LEP bound of $m_{\tilde\chi_1^0}>46$~GeV~\cite{Amsler:2008zzb} is based on chargino searches and their constraints on the combination of low values for $\mu-M_2$, 
assuming a fixed GUT relationship between the gaugino masses, $M_1\approx\frac{1}{2} M_2$. The invisible width of the $Z$ results in constraints of a similar significance on $\mu$ and $M_2$, and not directly on the neutralino mass. In fact, even a nearly massless neutralino is not completely excluded in the MSSM, however, a light neutralino must be dominantly bino in nature. For a recent brief review of the situation, see~\cite{Dreiner:2009yk}.

We have used Bayesian sampling techniques to explore a 24-dimensional parametrization of the MSSM, assuming diagonal soft supersymmetry breaking sfermion mass terms, independent soft mass terms for the gauginos, non-zero trilinear couplings for the third generation sfermions, and keeping $\mu$, $\tan\beta$ and $m_A$ as free parameters in the Higgs sector. We use the nested sampling Monte Carlo method developed by Skilling~\cite{Skilling}, with the algorithm implemented in the {\tt MultiNest} program~\cite{Feroz:2007kg,Feroz:2008xx}. For another use of multi-nested sampling in scanning the MSSM parameter space, and a more extensive discussion of the method, see~\cite{AbdusSalam:2009qd}. Our scan required approximately one week of continuous running on a single modern CPU, though we note that {\tt MultiNest} has a parallel implementation that can be used to generate results significantly quicker.

The likelihood we use for our scan is constructed from the predicted lightest neutralino mass and its velocity weighted total cross section, assuming independent measurements with central values and Gaussian widths of  $m_{\tilde\chi_1^0}=30\pm 5$~GeV and $\left< \sigma v \right>=(9.0\pm 0.9)\times 10^{-26}$~cm$^3$~s$^{-1}$. We do not use other observables directly in the likelihood, besides some direct mass constraints on sparticles (see below), as we wish to focus on the effect of the indirect detection observables. For our parameters we use flat priors with masses in the range of 10~GeV to 4~TeV, with the exception of the gluino mass parameter $M_3$ for which we allow $-4$~TeV to 4~TeV,  the trilinear couplings that vary from $-1000$ to 1000, and $\tan\beta$ which lies in the interval $2-60$. 

To evaluate the observables at each given MSSM parameter point investigated by {\tt MultiNest}, we calculate the sparticle spectrum using the {\tt  Isasusy} patch in {\tt Isajet 7.78}~\cite{Paige:2003mg}, assigning zero likelihood to points determined to be unphysical. The spectrum is then transferred to {\tt MicrOMEGAs 2.2}~\cite{Belanger:2001fz,Belanger:2004yn} which is used to calculate the resulting Dark Matter cross section and density.

The sampling output consists of posterior PDFs for the model parameters, sparticle masses and a few other quantities of interest, such as the Dark Matter density. In addition, we have implemented the model independent LEP direct search bounds on the slepton and chargino masses as found in the {\tt MicrOMEGAs} code, and assigned model points in violation a zero likelihood. The resulting $\tilde\chi_1^0$ mass and velocity averaged annihilation cross section have Gaussian distributions around the assumed central values, thus some {\it a priori} allowed models in the MSSM are clearly capable of accommodating the input measurements. Indeed, the point with highest likelihood exactly matches the input measurements, and has annihilation predominantly to $b\bar b$ pairs. We must now rationalise what our most likely models look like, and also determine whether they are consistent with other existing particle and astrophysical data.

\begin{figure*}[p!]
\begin{center}
\includegraphics[angle=0,width=0.90\textwidth]{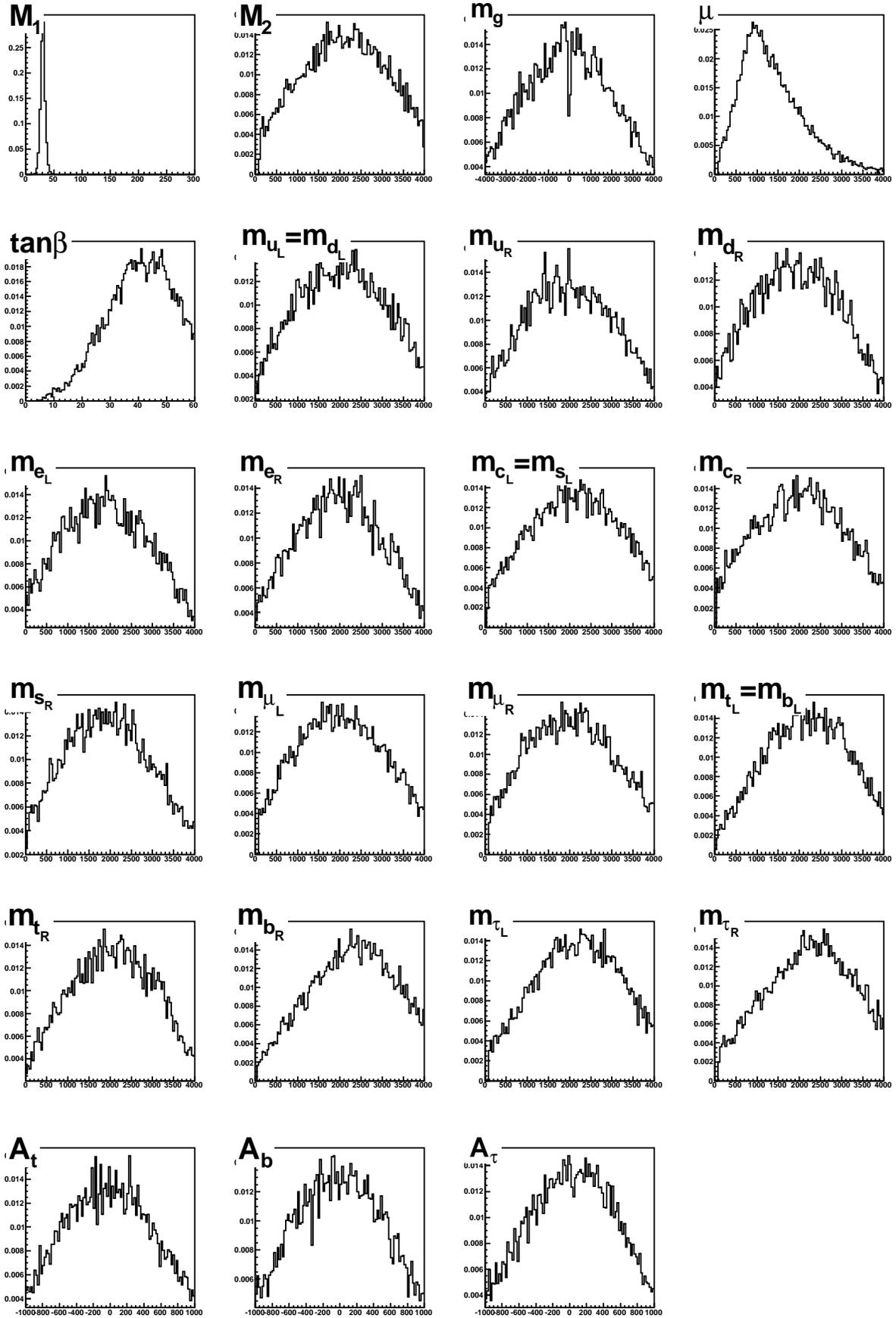}
\caption{Posterior PDFs for all weak scale input parameters.}
\label{fig:mssm}
\end{center}
\end{figure*}

\begin{figure}[ht]
\begin{center}
\includegraphics[angle=0,width=0.49\linewidth]{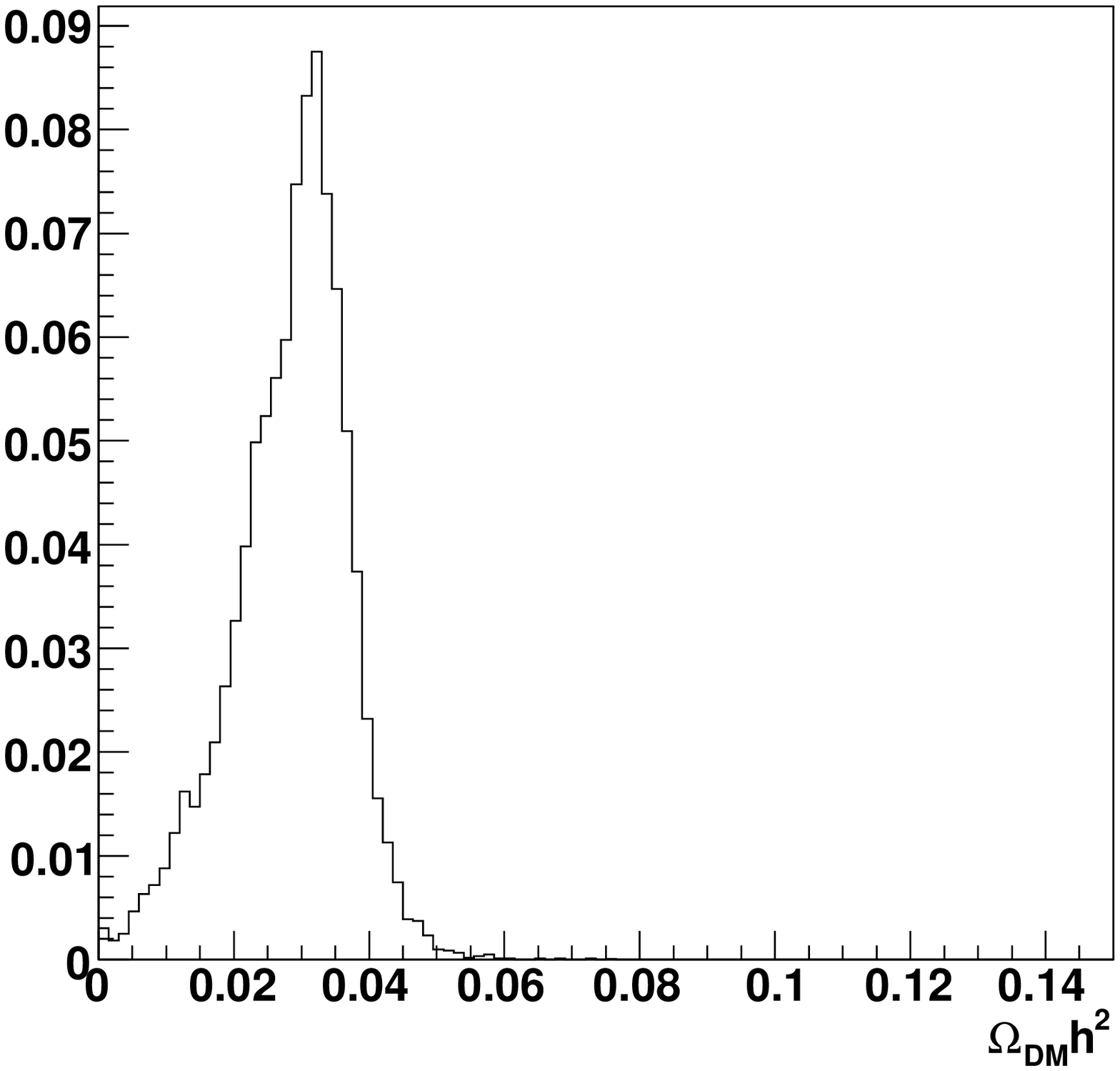}
\includegraphics[angle=0,width=0.49\linewidth]{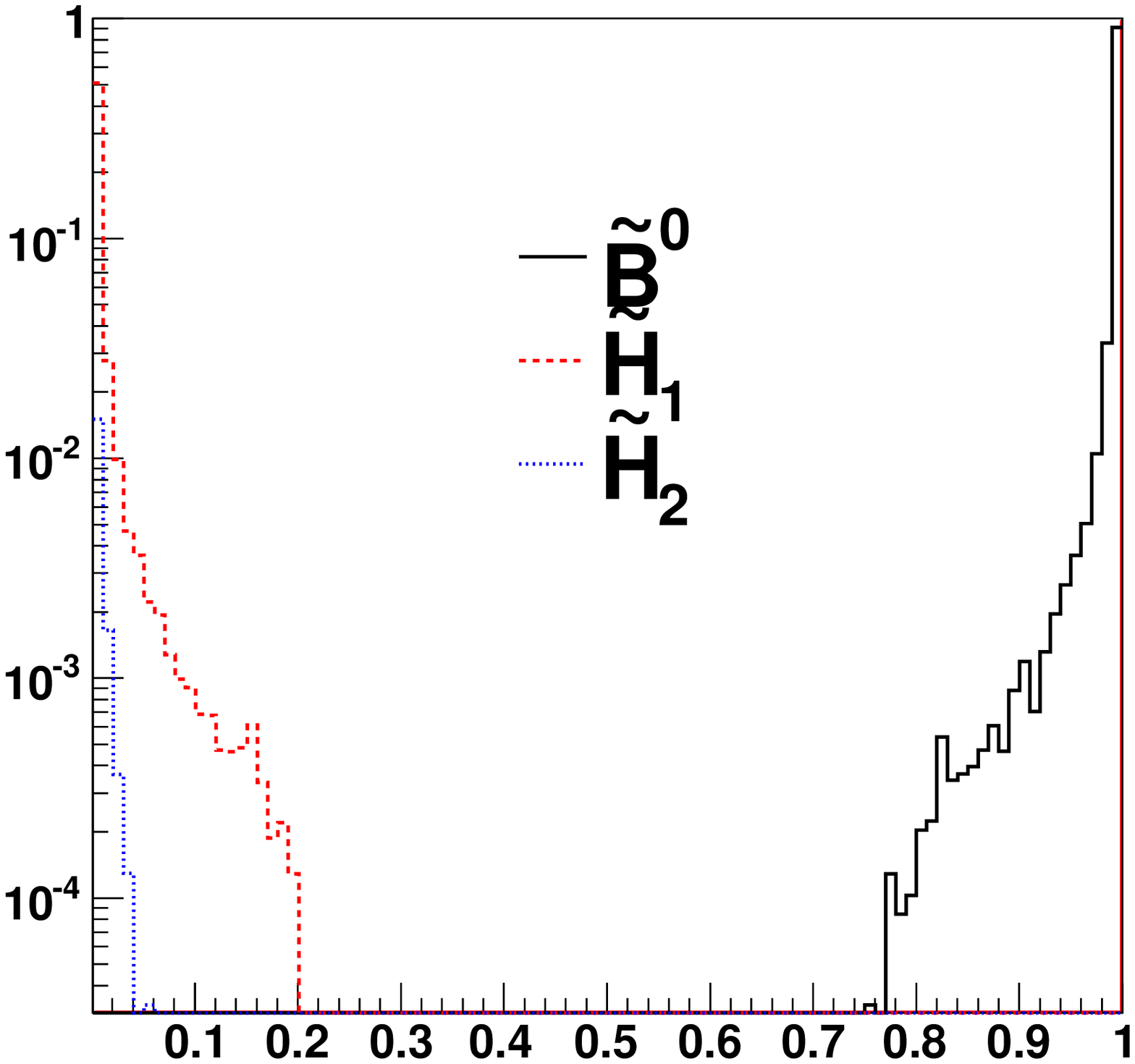}
\caption{Posterior PDF for Dark Matter density $\Omega h^2$ (left) and neutralino make-up (right).}
\label{fig:omegaDM}
\end{center}
\end{figure}

The marginalised posterior PDF of each weak scale input parameter is shown in Fig.~\ref{fig:mssm}. Of particular interest is: i) the tight constraint conferred on $M_1$, due to the equally tightly constrained $\tilde\chi_1^0$ mass and its bino nature, and ii)  the general preference for higher values of $\tan\beta$. As expected, the models with the greatest likelihood generally show a dominantly bino $\tilde\chi_1^0$ (see the right panel of Fig.~\ref{fig:omegaDM}), but with a small higgsino component. This is necessary in order to have a light neutralino that would have remained unseen at LEP, which can then only achieve the large annihilation cross-section through a Higgs resonance, using $2m_{\tilde\chi_1^0}\approx m_h,m_A$, preferring a small but non-zero higgsino component and large $\tan\beta$.

\begin{figure}[ht]
\begin{center}
\includegraphics[angle=0,width=.49\linewidth]{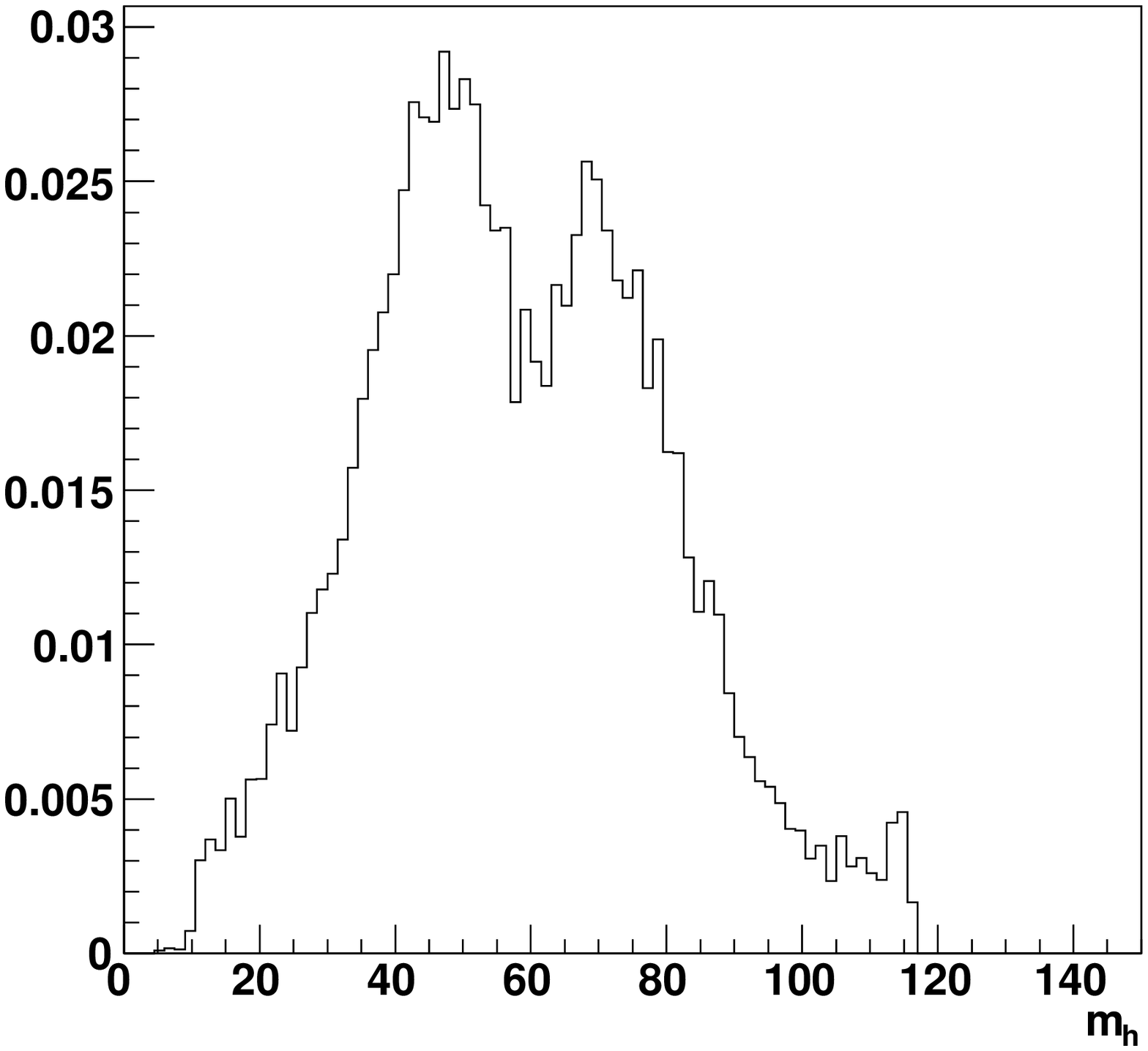}
\includegraphics[angle=0,width=.49\linewidth]{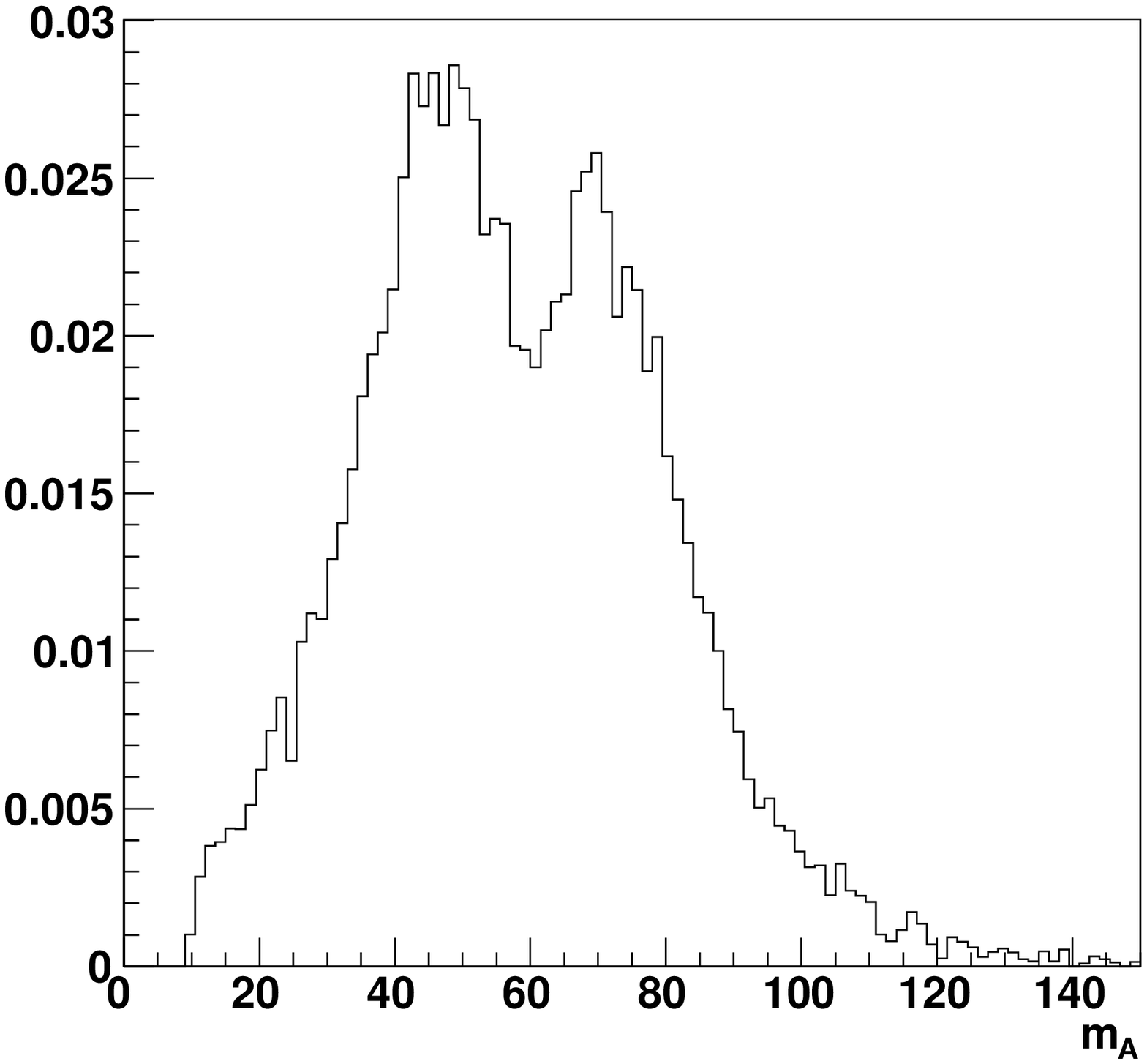}
\caption{Posterior PDFs for $m_h$ (left) and $m_A$ (right).}
\label{fig:higgsmass}
\end{center}
\end{figure}

We display the resulting Higgs mass PDFs in Fig.~\ref{fig:higgsmass}, in which one indeed observes a resonant annihilation peak occurring at roughly twice $m_{\tilde\chi_1^0}$, split into a double peak since the annihilation becomes too efficient when sitting exactly on the resonance, leading to a corresponding decrease in likelihood for these models. The Higgs masses are also fairly correlated, favouring $m_h \approx m_A \approx 2m_{\tilde\chi_1^0}$ for efficient annihilation. The cut off in the $m_h$ distribution at high Higgs masses comes from the upper limit on this mass, given the squark mass priors, whilst the small rise just before the cut off is the result of models in which the neutralino acquires a successively greater higgsino component pushing into higher neutralino masses and closer to the LEP limits on $\mu$.

It is evident from Fig.~\ref{fig:higgsmass} that the LEP bounds on the neutral MSSM Higgs masses are amongst the largest hurdles for an MSSM interpretation. Evaluating the precise limits given an arbitrary set of weak scale MSSM parameters is non-trivial, though one can get conservative limits by looking at the combined LEP exclusion bounds for the various extreme scenarios in~\cite{Schael:2006cr}. It is interesting to note that the weakest bounds at low masses generically occur when $m_h \approx m_A$ and $\tan\beta$ is relatively large, which is just what is favoured in our scan. This is due to the more difficult to reconstruct Higgs pair-production mechanism, $e^+e^-\to hA$, becoming kinematically allowed and competitive with Higgsstrahlung, $e^+e^-\to hZ$. As a result we take a very conservative limit of 85~GeV for both masses, allowing some points that may be excluded in a more rigorous analysis. Even so, this bound clearly excludes a large fraction of our models, including the most likely. 

Next, we consider the dark matter relic density PDF, shown in Fig.~\ref{fig:omegaDM} (left), which lies significantly lower than the value $\Omega h^2\simeq 0.1$ indicated by state-of-the-art $\Lambda$CDM fits. This is an inevitable consequence of imposing a reasonably high annihilation cross-section, but does not on its own exclude the models. Mechanisms for significant boosts in the density through modification of early universe cosmology have been suggested, see {\it e.g.}~\cite{Arbey:2008kv}, and the possibility of multi-component Dark Matter remains. Had the models given \emph{greater} relic densities relative to the existing data, one could be more confident in excluding them. We also show the PDFs for the neutralino components in Fig.~\ref{fig:omegaDM} (right), confirming its dominant bino nature, with some small higgsino admixture and effectively no wino component.

Finally, we consider the impact of the usual selection of precision observables, summarised in Table~\ref{table:precision}. Though it is difficult to make completely generic statements as to why a certain numbers of models fail a certain constraint, due to the number of parameters potentially affecting each observable, we offer general remarks below.

\begin{table}[ht]
\caption{Percentage of our posterior samples that fail cuts on precision observables. The observables were calculated using {\tt MicrOMEGAs}.}
\vspace{1mm}
\begin{tabular}{c|c|c}
Observable & Experimental Limit & Models excluded \\
\hline 
BR$(B \to X_s \gamma)$&$(3.55 \pm 0.42) \times 10^{-4}$~\cite{AbdusSalam:2009qd}&46.4\% \\
BR$(B_s \to \mu \mu$)&$<5.8 \times 10^{-8}$~\cite{:2007kv}&41.7\% \\
$\delta a_{\mu}$&$(30.2 \pm 9.2) \times 10^{-10}$~\cite{AbdusSalam:2009qd}&99.7\%\\
\end{tabular}
\label{table:precision}
\end{table}

The branching ratio for the decay $B \to X_s \gamma$ is tightly constrained experimentally to lie close to the SM value, but can be modified within the allowed range by SUSY contributions, predominantly involving loops with charged Higgs bosons and charginos. The chargino contribution is enhanced at large $\tan\beta$ and has potential large logarithms of $m_{\rm SUSY}/m_W$. For the allowed range we use the evaluation in~\cite{AbdusSalam:2009qd}. A significant number of our models give BR($B \to X_s \gamma$) values higher than the measured upper limit, in keeping with the observed preference for higher $\tan\beta$.

Similarly, the branching ratio for the decay $B_s \to \mu \mu$ can be enhanced in the MSSM via interactions involving neutral Higgs bosons. The effects are again higher at large $\tan\beta$, and one would expect an increase with lower neutral Higgs masses. A large number of our models are excluded by the the CDF Run II constraint on this branching ratio~\cite{:2007kv}, including our best fit model, and those that remain generally have lower values of $\tan\beta$.

By far the largest impact on our models comes from requiring the correction to the muon anomalous magnetic moment, $\delta a_{\mu}$, to account for the observed discrepancy with the SM. Almost all of our posterior points are inconsistent due to having too low $\delta a_{\mu}$ values. The main SUSY contributions to $a_{\mu}$ result from smuon-neutralino and sneutrino-chargino loops~\cite{Czarnecki:2001pv}. Our models generally prefer heavy neutralinos and charginos $\tilde\chi_i$, for $i>1$, due to a large $M_2$, pushed to higher values to maintain the specific character of the $\tilde\chi_1^0$ required by our input data, while the limit on the lightest chargino mass prevents $\mu$ from becoming very low. As a check we have taken our highest likelihood model, lowering the smuon mass parameters whilst also lowering $M_2$, which indeed increases $\delta a_{\mu}$.

If one discounts the apparent 3$\sigma$ discrepancy between the experimentally observed $a_{\mu}$ and the SM prediction, the posterior survives more or less unscathed. A thorough investigation of each of these precision observables would no doubt reveal further sensitivities and correlations but we considered this beyond the scope of this short Letter. 

Given the above, it is reasonable to ask if \emph{any} MSSM models consistent with all current experimental limits on supersymmetry remain compatible with the Goodenough and Hooper claim. Indeed 31 points, out of a total of 45301 posterior points, survive after imposing the bounds in Table~\ref{table:precision}, and in fact all have reasonable likelihoods, having annihilation cross-sections and neutralino masses close to the input values. These models have a significantly lower mass for the first and second generation sleptons, as one would expect from imposing the $\delta a_{\mu}$ constraint. The most likely point remaining after imposing the conservative LEP Higgs bound detailed above has $m_{DM}\simeq 31.8$~GeV, a velocity averaged annihilation cross section of $\left< \sigma v \right> \simeq 8.73\times 10^{-26}$~cm$^3$~s$^{-1}$ and Higgs masses $(m_h,m_A)=(96.8,96.5)$~GeV, and is thus not too far from satisfying the Goodenough and Hooper claim. There are a few models with Higgs masses greater than 110 GeV, but these are considerably less likely. If one ignores the lower bound on $\delta a_{\mu}$, the most likely model has $m_{DM}\simeq 30.3$~GeV and $\left< \sigma v \right> \simeq 9.04\times 10^{-26}$, and $(m_h,m_A)=(106.9,106.5)$~GeV. 

In conclusion, we have shown that the Dark Matter interpretation of the Fermi LAT data from the Galactic Centre in~\cite{Goodenough:2009gk} is generally hard to reconcile with neutralino Dark Matter in the MSSM. However, a small region of parameter space remains with a very light bino neutralino with some higgsino admixture, Higgs masses $m_h$ and $m_A$ near the border of the LEP excluded region, fairly light sleptons and moderate values of $\tan\beta$. This demonstrates the power of multi-modal nested sampling --- using only approximate information on the Dark Matter mass and annihilation cross section, as expected from a indirect detection signal --- to constrain models with a large number of parameters and complicated degeneracies, performing an efficient scan over a many-dimensional parameter space. 

Our current scan was very efficient in restricting the MSSM parameter space due to the difficulty in the MSSM of simultaneously getting a low neutralino mass and a large annihilation cross section, while at the same time evading earlier collider bounds. We therefore feel that one would do well to repeat this interesting case study if more definite evidence of Dark Matter is found in indirect detection experiments, taking care to also incorporate properly the astrophysical uncertainties associated with the halo distribution. Finally, although it is possible to get close to accommodating the Goodenough and Hooper claim for a small fraction of our models, dedicated fans of neutralino Dark Matter in the MSSM should probably hope that the present interpretation is a false alarm.

{\it Acknowledgements:} ARR and MJW thank the members of the Cambridge Supersymmetry Working Group and the Oscar Klein Centre for helpful discussions. MJW also thanks Pamela Ferrari for useful conversations regarding LEP Higgs constraints. MJW acknowledges funding from the UK Science and Technology Facilities Council (STFC) and ARR is grateful for financial support from the Swedish Research Council (VR) through the Oskar Klein Centre.

\end{document}